\documentstyle[12pt]{article}

\makeatletter
\@addtoreset{equation}{section}
\makeatother

\def\RR{{{\rm l}\kern-.17em{\rm R}}}
\newcommand\bfab{\em}
\newcommand\Gh{{\hat{G}}}

\newcommand\sbr{\overline{s}}
\newcommand\wbr{{\overline{w}}}
\newcommand\Dub{{\overline{\Delta}_N}}
\newcommand\Dlb{{\underline{\Delta}_N}}

\newtheorem{thm}{Theorem}[section]
\newtheorem{lemma}[thm]{Lemma} %[section]
\newtheorem{cor}[thm]{Corollary} %[section]
\newcommand{\EQ}{\begin{equation}}
\newcommand{\EN}{\end{equation}}
\newcommand{\BEQ}{\begin{equation}}
\newcommand{\NEQ}{\end{equation}}
\newcommand{\nn}{\nonumber}

\def\ni{\noindent}

\def\eopp{{$\Box$}}      %\vrule height7pt width7pt depth0pt} }%\par\bigskip}
\def\xbf{{\bf x}\,}
\def\Ebf{{\bf E}\,}

\def\ftl{\tilde{f}}
\def\fh{\hat{f}}

\def\fhl{\hat{f}^{(\ell)}}

\def\fhlN{\hat{f}_N^{(\ell)}}
\def\fhlpN{\hat{f}_N^{(\ell+1)}}
\def\eps{{\epsilon}}

\def\part{\partial}

\date{ }

\begin{document}
\bibliographystyle{plain}

\title{Piecewise Convex Function Estimation: \\ %Robustness, 
Representations, Duality
and Model Selection}
%The challenge of Geometric Fidelity}
%\def\chptitle{Piecewise Convex Model Selection} 

\author{Kurt S.~Riedel \\
Courant Institute of Mathematical Sciences \\
New York University \\
New York, New York 10012-1185}

\maketitle
\begin{abstract} %\abstract
{We consider spline estimates which preserve prescribed piecewise
convex properties of the unknown function.
A robust version of the penalized likelihood is given and shown to
correspond to a  variable halfwidth kernel smoother where the halfwidth 
adaptively decreases in regions of rapid change of the unknown
function. When the convexity change points are prescribed,
we derive representation results and smoothness properties of the
estimates. A dual formulation is given which reduces
 the estimate is reduced to a finite dimensional convex
optimization in the dual space. 
 %Given noisy data, function estimation is considered when the unknown
%function is known {\it a priori} to consist of a small number of regions
%where the function is either convex or concave. If the regions are known
%{\it a priori},
}\end{abstract}

%{\vskip-.40in}

\section{Introduction}\label{I}

A common problem in nonparametric function estimation is that the
estimate often has artificial wiggles
that the original function does not possess. 
In practice, these extra inflection points
have a very negative impact on the utility and credibility of
the estimate.  
%Our basic tenet is:
%``piecewise $\ell$-convex with a small number of
%change points of convexity.''
In this article, we examine function estimation which preserves
the geometric shape of the unknown function, $f(t)$. 
%We are given $N$ noisy measurements of the unknown function, $f(t)$, 
%contaminated with random noise, 
%and we seek to estimate $f(t)$ while preserving the geometric
%fidelity of the estimate, $\fh(t)$, with respect to the true function. 
In other words, the number and location of the change points of convexity of
the estimate, $\hat{f}(t)$, should approximate those of ${f}(t)$.

%ore generally, we can impose piecewise $\ell$-convexity in place of
%onvexity. $\ell$-convexity means that the $\ell$-th derivative of $f(t)$
%s nonnegative, $f^{(\ell )} (t)\geq 0$.
We say
that $f(t) $ has $K$ change points of $\ell$-convexity  with change points
$x_1 \leq x_2 \ldots \leq x_K$ if $(-1)^{k-1} f^{(\ell )} (t) \geq 0$ for
$x_k \leq t \leq x_{k+1}$.  
%with strong sign changes of $f^{(\ell )} (t)$ at $x_k$.
For $\ell = 0$, $f(t)$ is
nonnegative and for $\ell =1$, the function is nondecreasing. In regions
where the constraint of $\ell$-convexity is active, $f^{(\ell )} (t) = 0$ and
$f(t)$ is a polynomial of degree $\ell -1$. 
For 1-convexity, $f(t)$ is constant in
the active constraint regions and for 2-convexity, the function is linear.
Our subjective belief is that most people prefer smoothly varying functions
such as quadratic or cubic polynomials even in the active constraint regions.
Thus, piecewise 3-convexity or 4-convexity are also reasonable hypotheses.

%The idea of constraining the function fit to preserve 
When the change points are {\em prescribed}, convex analysis can
be employed to derive representation theorems, smoothness properties
and duality results.
%$\ell$-convexity properties has been considered by a number of authors
Our work extends that of 
Refs.~\cite{Deutsch,Maechler,Michelli,MU85,MU91,Utreras,VW} 
to an important class of robust functionals.
To motivate robust nonparametric estimation,
we show that robust functionals correspond to variable
halfwidth data-adaptive kernels, i.e.   the effective halfwidth decreases
in regions where the unknown function changes rapidly.
%The more
%difficult problems of determining the number and location of the
%$\ell$-convexity breakpoints is the focus of  \cite{Li95,Mammen91,Riedel2}. 
When the number of change points are known, we prove the existence
of a minimum of the penalized likelihood.
 
%We refer to the estimation of the number of change points as the
%``model selection problem'' because it resembles model selection in an
%infinite family of parametric models.

Sections \ref{FAP} and \ref{CCC} contain functional analysis preliminaries.
Lemma \ref{Lcone} gives a characterization of negative polar cones
in the spirit of \cite{Deutsch}. 
Representation and duality theorems for constrained smoothing splines
have been developed in 
\cite{MU85,MU91,Utreras}
for the case of prescribed convexity change points.
In Sections \ref{SMSPL} and \ref{DUAL},
we generalize these results to the case of nonquadratic penalty functions.
In Section \ref{ROBUST}, we show how robust penalty functions
correspond to data-adaptive variable halfwidth kernel smoothers. 
In Section \ref{CPsect}, 
we consider estimating the change point locations by minimizing
the penalized least squares fit.
%In Section \ref{MODEL}, we consider adaptive knot placement and
%propose a  model selection criterion which penalizes for extra inflection
%points.

\section{Functional Analysis Preliminaries}\label{FAP}
%{\vskip-.1in}

We consider an unknown function, $f$, is in the Sobolev space
$W_{m,p} [0,1]$ with $m\geq \ell$ and $1<p< \infty$, where
\begin{equation}\label{2.1}
%{\vskip-.22in}
W_{m,p} \equiv\ \{  f|f^{(m)}  \in L_p [0,1] \ {\rm and\ } f,f^{\prime} \ldots
f^{(m-1)} (t) \ {\rm  \ absolutely \ continuous} \} \ .
%\eqno (2.1)
%{\vskip-.1in}
\end{equation}
\noindent
For $f\in W_{m,p}$, we have the representation
\begin{equation}\label{2.2}
f(t) = \sum_{j=0}^{m-1} a_j P_j (t) + \int_0^t {(t-s)^{m-1} \over
(m-1)!} f^{(m)} (s) ds \ ,
%\eqno ({2.2})
\end{equation}%\label{}
where $P_j (t) \equiv t^j /(j!)$.
Equation (\ref{2.2}) decomposes $W_{m ,p}$
into a direct sum of the space of polynomials of degree $m-1$  %, $P_{m-1}$
plus the set of functions whose first $m-1$ derivatives vanish at $t=0$,
which we denote by $W_{m,p}^0$ \cite{Wahba91}. 
We define the seminorm
$\|f\|^p_{j,p} \equiv \int_0^1 |f^{(j)} (t) |^p dt$.
%when the subscript is absent, the seminorm is $\|\cdot \|_{0,2}$.
We endow $W_{m,p}$ with the  norm:
\begin{equation}\label{2.3}
%\vskip-.15in} 
|\| f \|\vert^p_{m,p} = \sum_{j=0}^{m-1} |f^{(j)} (t=0) |^p +
\|f\|^p_{m,p}         %\int_0^1 |f^{(m)} (t) |^p dt 
\ .
%\eqno (2.3) 
\end{equation}
%{\vskip-.1in}
\noindent 
The dual space of $W_{m,p}$ is isomorphic to the direct sum 
of $P_{m-1}$ and $W_{m,q}^0$ with $q=p/(p-1)$ and 
the duality pairing of $f\in W_{m,p}$ and  $g \in W_{m,q}$ is 
%By the Riesz representation theorem, every element,
%$B$, in the dual space, $W_{m,p}^*$ has the representation
%{\vskip-.15in}
\begin{equation}\label{2.4}
\langle\langle g,f\rangle\rangle =
\sum_{j=0}^{m-1} b_j a_j + \int_0^1 f^{(m)} (t) g^{(m)} (t) dt 
\ ,  %\eqno (2.1)
\end{equation}
%{\vskip-.1in}\noindent
%In (\ref{2.4}), 
where $a_j \equiv f^{(j)}(0)$ 
and $b_j \equiv g^{(j)}(0)$. 
We denote the
duality pairing by $\langle\langle \cdot\rangle\rangle $  and the $L_2$ inner
product by $\langle \cdot\rangle $.
%For the duality results of Section III, we will need the reproducing kernel
%formulation of $W_{m ,p}$.
In \cite{Wahba91}, $W_{m,2}$ is given a reproducing kernel  where
for each $t$, $f(t) = \langle \langle R_t ,f\rangle\rangle$.
The same reproducing kernel structure carries over to $1<p<\infty$ with
%The space $W_{m ,p}$
%has a reproducing kernel, $R(t,s)$, 
%The reproducing kernel structure of $W_{m,2}$ is described in \cite{Wahba91}.
%such that for each $t$, $f(t) =\langle \langle R_t ,f\rangle\rangle$ 
%\cite{Wahba91}.
\begin{equation}\label{2.5}
R_t (s) = \sum_{j=0}^{m-1} P_j (t)P_j (s) + \int_0^{\min \{ t,s\} }
{(t-u)^{m-1} (s-u)^{m-1} du \over [(m-1)!]^2 } \ .
\end{equation}
%The natural boundary conditions are automatically found: 
%$\partial_s^{m+k} R_t (s) = 0$, for $s=0,1$ and $k=0\ldots m-1$.
%Note $\partial_s^m R_t (s) = {(t-s)^{m-1}_t \over (m-1)!}$ so for
%$R_t \in W_{m,p}^0$,
%\begin{equation}\label{ad}
%f(t) =\langle\langle R_t f\rangle\rangle =
%\int_0^1 \partial_s^m R_t (s) f^{(m)} (s)ds
%= \int_0^1 {(t-s)^{m-1} \over (m-1)!} f^{(m)} (s) ds 
%\end{equation}%\label{}
A linear operator $L_i \in
W_{m,p}^*$ has representations 
$L_i f = \langle\langle b_i(s),f(s)\rangle\rangle$ and
$L_i f = \langle h_i ( s),f(s)\rangle$, 
where  $b_i(s) \equiv  L_i R(\cdot ,s)$ and  
$h_i(s) \equiv  L_i \delta (s-\cdot )$.
$L_i R(\cdot ,s)$ denotes $L_i$ acting on the first entry of $R$.
In the standard case, where
$L_if \equiv f(t_i )$, $b_i(s) = R(t_i,s)$ and $h_i(s) =\delta(s-t_i)$.
%where $R_t (s)$ is the reproducing kernel of (\ref{2.5}).
%We represent $L_if$ as and assume $h_i \in W_{\ell,1}^{\perp}$. 
%For example if $L_i f=\partial_t^2 f(t_i )$, $L_i$ has the kernel
%representation $\partial_t^2 R(t_i ,s)$.

%We represent $L_if$ as $b_i(s) \equiv  L_i R(\cdot ,s)$ and  
%$h_i(s) \equiv  L_i \delta (s-\cdot )$ and assume 
%$h_i \in W_{\ell,1}^{\perp}$. In the standard case where
%$y_i = f(t_i ) + \epsilon_i$, $b_i(s) = R(t_i,s)$ and $h_i(s) =\delta(s-t_i)$,
%where $R_t (s)$ is the reproducing kernel of (2. ).

\section{Convex Cone Constraints}\label{CCC}

In this and the next section, 
we assume that the change points $\{ x_1 \ldots x_K \}$
of $\ell$-convexity are given and that the unknown 
function is in the Sobolev space, $W_{m,p} [0,1]$.
Given change points, $\{ x_1 ,x_2 \ldots x_K \}$, we define the closed
convex cone
%{\vskip-.2in}
\begin{equation}\label{3.1i}
V^{K,\ell }_{m,p} [x_1 ,\ldots ,x_K ] = \{ f \in
W_{m,p}  \ | \ (-1)^{k-1}
f^{(\ell )} (t) \geq 0 \ \ {\rm for } \ \ x_{k-1} \leq t \leq x_k \} \ ,
\end{equation}
where $x_0 \equiv 0$ and $x_{K+1} \equiv 1$.
%{\vskip-.1in}
Let $\bf{x}$  denote the $K$ row vector, $( x_1 ,x_2 \ldots x_K )$.
Throughout this article, we require     $\ell \le m$. 
By the Sobolev embedding theorem, $f^{(\ell )} (t)$ is continuous
for $\ell < m$. For $\ell=m$, we require the convexity constraint
in (\ref{3.1i}) almost everywhere.
%The negative polar of $V^{k,m }_{m,p} [\xbf]$ is
%$V^{k,m}_{m,p} [\xbf]^- = -V^{K,\ell }_{m,p} [\xbf] \cap W_{m,p}^0$. 
We define the class of functions with at most $K$ change points as
%{\vskip-.2in}
\begin{equation}\label{3.2i}
V_{m,p}^{K,\ell} \equiv
\bigcup_{x_1 \leq x_2 \ldots \leq x_K }
\bigl\{ V^{K,\ell}_{m,p} [x_1 ,\ldots ,x_K ] \cup (
-V^{K,\ell}_{m,p} [x_1 ,\ldots ,x_K ] ) \bigr\} \ .
\end{equation}
%{\vskip-.15in}
By allowing $x_{k^{\prime} }
= x_{k^{\prime} +1} $, we have embedded
$V^{K,\ell}_{m,p}$ into
$V^{K+1,\ell}_{m,p}$.
%Note that
$V^{K,\ell}_{m,p}$
is the union of convex cones, and is closed but not convex. 
Similar piecewise $\ell$-convex classes are
defined in \cite{Mammen91} for the case $\ell=m+1$ with 
a supremum norm on the H\"older constant for $f^{(\ell-2)}$.
%To  decompose $W_{m,p}$ in terms of $V^{K,\ell}_{m,p}$,
%we require that each function in $W_{m,p}$
%has a piecewise continuous $\ell$-th derivative.
%By the Sobolev embedding theorem,
%this corresponds to the case $m\geq \ell +1$.

For Theorem \ref{Dthm}, we need the following results from convex analysis.

\ni
{\bf Definition} [1] {\em  Let $C$ be a closed convex cone in $W_{m,p}$;
the negative polar, $C^-$, of $C$ is
$C^- \equiv \{g \in   W_{m,p}^* |\ \forall f \in C,
\langle\langle g,f \rangle \rangle \le 0 \}$. 
}

For  $V^{K,\ell}_{m,p} [\xbf]$, we are able to give a more explicit  
characterization of the negative polar. Our  result   is restricted to 
$\ell \le m$ while Deutsch et al. \cite{Deutsch} 
consider the more difficult case
of $m=0$ with $\ell \ge m$.

\noindent
\begin{lemma} \label{Lcone}
{  
The negative polar of $V^{K,\ell}_{m,p} [\xbf]$ is
$V^{K,\ell }_{m,p} [\xbf]^- =$ closure in $W_{m,p}^*$ of
the $\{ g\ \in C^{2m-\ell}[0,1]\ |\ %{\rm st}\  
(-1)^{m-\ell}
\chi_{\bf x}(t)g^{(2m-\ell)}(t) \le 0; \ \ 
g^{(j)}(0)=0,\ j = 0 \ldots \ell-1;\ g^{(m+j)}(1)=0,\ {\rm and} \ \
g^{(m-j-1)}(0) -(-1)^{j} g^{(m+j)}(0) = 0; \ \ 
 j = 0 \ldots m - \ell -2;
(-1)^{m-\ell}\chi_{\bf x}(1)g^{(2m-\ell-1)}(1)$ 
$\ge 0;\ 
\chi_{\bf x}(0)[g^{(\ell)}(0)+ (-1)^{m-\ell}
g^{(2m-\ell-1)}(1) \le 0 \}$, 
%{\rm for}\ f \in V^{k,m- 2}_{m,p} [\xbf] $
where $\chi_{\bf x}(t)$ is defined by 
$\chi_{\bf x}(t) = (-1)^{k-1}$
{\rm for } $ x_{k} < t < x_{k+1} \}$ 
and $\chi_{\bf x}(x_k)=0,\ \ k = 1 \ldots K$. } 
\end{lemma}

\noindent
{\bfab Proof.} 
Integration by parts yields
$\int_0^1 f^{(m)} (t) g^{(m)} (t) dt = (-1)^{m-\ell}
\int_0^1 f^{(\ell)} (t) g^{(2m-\ell)} (t) dt\ + $ 
$\sum_{j= 0}^{m-\ell-1}(-1)^{j}  f^{(m-j-1)} (t) g^{(m+j)}|^1_0$
for $g \in C^{2m-\ell}[0,1]$. We now find test functions, 
$\tilde{f}\in V^{K,\ell}_{m,p} [\xbf]$ which require each term separately to be
nonpositive. For $t \ne x_K$, we choose $\tilde{f}^{(\ell)}(s)=
(s-t+h)_+^{m-\ell+1} (t-s+h)_+^{m-\ell+1} \chi_{\bf x}(s)$
where $h$ is a small localization parameter and $s_+ \equiv s$ for $s>0$
and zero otherwise. The boundary conditions at $t=1$
are proved inductively with
the sequence of test functions $\tilde{f}_h(s)=
(s-1+h)_+^{m} \chi_{\bf x}(s)$ as $h\rightarrow 0$.
\eopp

\begin{cor}
%{\bf Corollary 1.}
{ %\em
The negative polar of $V^{K,m }_{m,p} [\xbf]$ is
$V^{K,m }_{m,p} [\xbf]^- = -V^{K,m }_{m,p} [\xbf] \cap W_{m,p}^0$. 
}
\end{cor}

\begin{cor}
%{\bf Corollary 2.}
{ %\em
The negative polar of $V^{K,m-1}_{m,p} [\xbf]$ is the 
closure in $W_{m,p}^*$ of the $\{ g\ \in C^{m+1}[0,1]\ |\ 
g^{(m+1)}(t)\chi_{\bf x}(t) \ge 0;\ \chi_{\bf x}(1)g^{(m)}(1) \le 0;\  
\chi_{\bf x}(0)[g^{(m-1)}(0) -g^{(m)}(0)] \le 0\ \}
$. 
}
\end{cor}

\begin{cor}
The negative polar of $V^{K,m-2}_{m,p} [\xbf]$ is
$V^{K,m-2 }_{m,p} [\xbf]^- =$ closure in $W_{m,p}^*$ of
the $\{ g\  \in C^{m+2}[0,1]\ |\  g^{(m+2)}(t)\chi_{\bf x}(t) \le
0;\ g^{(m)}(1) =0;\  \chi_{\bf x}(1)g^{(m+1)}(1) \ge 0;\
[g^{(m-1)}(0) -g^{(m)}(0)] = 0;\  
\chi_{\bf x}(0)[g^{(m-2)}(0) +g^{(m+1)}(0)] \le 0 \}.$
%g^{(iv)}(t)f(t) \le 0, g''(1)=0, g'(0) -g''(0) = 0,
%f(1)g'''(1) \ge 0, f(0)[g(0)+g'''(1)] \le 0 $
\end{cor}

The negative polar is useful in evaluating the normal cone of $K$:
%as given by the following result.

\begin{lemma}[\cite{Aubin84}, p.171] \label{LemN5}
% {\bf Lemma 1.} 
{Let $C$ be a closed convex cone in W, 
the normal cone of $C$ in $W^*$ at $f$, $N_C(f)$ satisfies 
$N_C(f)= C^- \cap \{ f \}^{\bot}$, where $C^-$ is the negative polar of $C$.}
\end{lemma}

{\vskip-.2in}
\section{Robust splines: Representations and Smoothness} \label{SMSPL}
{\vskip-.12in}

%\noindent{\bf III. CONSTRAINED SMOOTHING SPLINES AND DUALITY}

In this section, we generalize representation and smoothness results
to a large class of robust functionals.
These robust functionals are advantageous because they downweight outliers
and adaptively adjust the effective smoothing width. 
We are given $N$ measurements of the unknown function, $f(t)$:
%{\vskip-.1in}
\begin{equation}\label{M1}
 \ y_i = L_if  + \epsilon_i =
\langle h_i ,f \rangle + \epsilon_i  =
\langle\langle b_i ,f \rangle\rangle + \epsilon_i \ , %\eqno(2.2)
\end{equation}
%{\vskip-.1in} \noindent
where the $L_i $ are bounded linear operators 
on $W_{m,p}$, % \ cap C^{-\ell}$ ,  
and the $\epsilon_i$ are %independent, normally distributed 
uncorrelated random variables with variance $\sigma_i^2 > 0$.
A robustified estimate of $f(t)$ given the measurements $\{y_i\}$ 
is $\hat{f} \equiv{\rm argmin}$ ${\rm{VP}}
[f\in V^{K,\ell}_{m,p} [x_1 ,\ldots ,x_K ] ]$:
\begin{equation}\label{3.1}
%{\vskip-.15in}
%hat{f} = {\rm arg \ min} 
{\rm VP}[f ] \equiv
% \in V^{K,\ell}_{m,p} [x_1 ,\ldots ,x_K ] ] \equiv  %\left\{ 
\frac{\lambda}{p}\int |f^{(m)} (s) |^p ds +
 \sum_{i=1}^N \rho_i \left( \langle h_i ,f \rangle -y_i \right) %\right\}
\ , \ \
%g \in V{K,\ell}_{m,p} [x_1 ,\ldots ,x_K ]%\eqno(2.3)
\end{equation}
%{\vskip-.12in}
\noindent
where the $\rho_i$ are strictly convex, continuous functions. The
standard case is $p=2$ and $\rho_i (y_i -\langle h_i ,f\rangle ) =
|y_i -f(t_i )|^2 /(N\sigma_i^2)$. For an excellent discussion of the
advantages of robustness in function estimation, see 
M\"achler \cite{Maechler}.
%\begin{equation}\label{3.21}
%{\rm VP}[f ] \equiv \frac{\lambda}{2}\int |f^{(m)} (s) |^2 ds + \frac{1}{N} 
%\sum_{i=1}^N \frac{|\langle h_i ,f \rangle -y_i|^2}{\sigma_i^2} %\right\}
%\ , \ \ \end{equation}

Theorem \ref{thm1} is proven in \cite{Utreras} 
and Theorem \ref{Dthm} is proven in \cite{MU85}
for the case $p=2$ and $\rho (y) = y^2$.
For the unconstrained case of Theorem \ref{thm1}, see \cite{Maechler}.
Equation (\ref{2.5}) and the corresponding smoothness results appear in 
\cite{Utreras}
for the case $\ell=1$, $p=2$ and $L_i = \delta(t-t_i)$.
The set of $\{ h_i,\ i=1,\ldots ,N \}$ separate polynomials of degree
$m-1$ means that  $\langle h_i , \sum_{k=0}^{m-1} c_k t^k \rangle
= 0$, $\forall i$ implies $c_k \equiv 0$. %Any $m$ distinct points,
%$t_i$, separate polynomials of degree $m-1$.

\noindent
\begin{thm} \label{thm1}
%{\bf Theorem 1.}
{Let $\{ h_i \}$ separate
polynomials of degree $m-1$; then the {mini}mization problem (\ref{3.1}) has
an unique solution in $V^{K,\ell}_{m,p} [\xbf]$, and  the
minimizing function %is in $C^{2m-\ell-2}$ and
satisfies the differential equation:
%{\vskip-.92in}
\begin{equation}\label{3.2}
(-1)^m \lambda d^m [|\fh^{(m)} |^{p-2} \fh^{(m)} (t) ] + 
\sum_{i=1}^N \rho_i^{\prime}
(\langle h_i ,\fh \rangle -y_i ) h_i(t) = 0 \ ,
%eqno (2.4)
\end{equation}
%{\vskip-.15in} 
\noindent
in those regions where $|f^{(\ell )} |>0$ for  $1 < p < \infty$
and $\ell \le m$.}
\end{thm}

\noindent
{\bfab Proof.} The functional (\ref{3.1}) is strictly convex, 
lower semicontinuous
and coercive, so by Theorem 2.1.2 of Ekeland and Temam \cite{Ekeland},
it has a unique minimum, $\fh$, on any closed convex set. 
From the generalized calculus of convex analysis, the solution satisfies
%{\vskip-.18in}
\begin{equation}\label{3.3D}
0 \in (-1)^m \lambda d^m [|\fh^{(m)} |^{p-2} \fh^{(m)} (t)]  \ + \
\sum_{i=1}^N \rho_i^{\prime} (\langle h_i ,\fh \rangle  -y_i ) h_i(t) +
\partial N_V(f)  \ ,       %\chi_{\bf x} (V^{K,\ell}_{m,p}[\xbf])(f) 
%\eqno(2.4)
\end{equation}       %{\vskip-.1in} 
\noindent
where $N_V(f)$ is the normal cone of  $V^{K,\ell}_{m,p} [\xbf]$ at $f$
\cite[p.~189]{Aubin84}.    %is the indicator function of $V^{K,\ell}_{m,p}$.
The normal cone is characterized by Lemmas \ref{Lcone} and \ref{LemN5}.
From \cite{Utreras}, each element of $N_V(f)$ is the limit of a discrete sum:
$\sum_{t} a_t \delta^{(\ell)}(\cdot-t)$ where the $t's$ are in the active
constraint region. Integrating (\ref{2.4}) yields
%{\vskip-.18in}
\begin{eqnarray} \label{3.4I}    %\eqalignno{ 
\lambda |\fh^{(m)} |^{p-2} \fh^{(m)} (t)  &=
\sum_{i=1}^N { \rho_i^{\prime} (\langle h_i ,\fh \rangle  -y_i )
\langle h_i(s), (s-t)_+^{m-1}\rangle \over (m-1)!}  \nn \\
 &
+ \int {(s-t)_+^{m-\ell-1}d\mu(s) \over  (m-\ell-1)!} \  ,
 %\chi_{\bf x} (V^{K,\ell}_{m,p}[\xbf])(f) 
\end{eqnarray}%{\vskip-.1in} 
\noindent
where $d\mu$ corresponds to a particular element of  $N_V(f)$.
\eopp

For $h_i(s) = \delta (s- t_i)$ and $\ell = m$, 
Theorem \ref{thm1} can be derived as a
consequence of the corresponding result for constrained interpolation
\cite{Michelli}.
\begin{cor}
%\noindent{\bf Corollary 2.}
{If $\{ h_i \}$ are in $W_{\ell,1}^{\perp}$, then the 
%polynomials of degree $m-1$, then the {mini}mization problem 
%an unique solution in $V^{K,\ell}_{m,p} [\xbf]$ and  the
minimizing function of (\ref{3.1}) is in $C^{2m-\ell-2}[0,1]$.
}
\end{cor}

\noindent
{\bfab Proof.}
Since $ (s-t)_+^{m-\ell-1}$ is $m-\ell-2$ times differentiable, 
the first term on the right hand side of 
(\ref{3.4I}) is $m-\ell-2$ times differentiable.
By hypothesis, $h_i \in  W_{\ell,1}^{\perp}$ and thus
$\int h_i(s)(s-t)_+^{m-1} ds$ is in  $C^{m-\ell-2}$.
Integrating (\ref{3.4I}) yields $f\in C^{2m-\ell-2}$.
\eopp

%For $V= V^{K,\ell}_{m,p}[\xbf]$, $N_V(f) =\{ g \in
%-V^{K,\ell }_{m,p} [\xbf] \cap W_{m,p}^0 | f^{(\ell )}
%(t) g(t) \equiv 0\}$. Since $|f^{(\ell )} |>0$ implies $N_V(f)(t)=0$,
%(\ref{2.4) follows. \eopp

%By Theorem --- of Aubin and Ekeland, $\partial \chi_{\bf x} (V^{K,\ell}_{m,p})(f)$
%consist of the set of functions, $g(t)$, such that $f^{(\ell )}
%(t) g(t) \equiv 0$ and  $g(t) \in V^{K,\ell}_{m,q}$.
%box
%\begin{enumerate}

\section{Equivalent adaptive halfwidth of robust splines}
\label{ROBUST}

Replacing the standard spline likelihood functional ($p=2$ and 
$\rho(y)= y^2/\sigma^2$) in (\ref{3.1})
with a more robust version 
has several well-known advantages.
First, outliers are less influential when $\rho(y)$ 
downweights large values of the residual error. 
Second, for $1\le p < 2$, the set of candidate functions
are increased, and the solution may have sharper local variation
than in the $p=2$ case.
We now describe a third important advantage: the effective halfwidth
adapts to the unknown function.

In \cite{Silverman84}, it is shown that as the number of measurements
increase the spline estimate converges to a local kernel smoother
estimate (provided the measurement times, $\{t_i \}$, are nearly regularly 
spaced.) For technical details, see   \cite{Silverman84}. 
%Appendix B of \cite{Riedel2}. 
Convergence proofs are available only
for $p=2$. The resulting effective kernel halfwidth, $h_{eff}$, is
scales as $h_{eff} \sim [\lambda F'(t) ]^{1\over 2m}$,
where $F(t)$ is the limiting distribution of measurement points.

For $1 < p < 2$, no theory exists on the effective halfwidth of
a robust spline estimate. 
We assume that $\fh$ converges to $f$ in $W_{m,p}$ under 
hypotheses similar to those used for the $p=2$ case in \cite{Silverman84}. 
%Appendix B of \cite{Riedel2}. 
These conditions relate to 
the discrepancy of the measurement times, $\{t_i\}$, the smoothness of $F(t)$,
and the scaling of the smoothing parameter with $N$.
The appropriate modifications for $p\ne 2$ are unknown.

We can make a heuristic two-scale 
analysis of (\ref{3.3D}) in the continuum.  
We assume that in the continuum limit, the estimate satisfies
the following equation to zeroth order:
%spline estimate converges to a kernel smoother with halfwidth, $h_{eff}(t)$. 
\begin{equation}\label{3.3C}
(-1)^m \lambda d^m [|\fh^{(m)} |^{p-2} \fh^{(m)} (t)]  \ + \
\fh(t) = y(t)  \ ,       %\chi_{\bf x} (V^{K,\ell}_{m,p}[\xbf])(f) 
%\eqno(2.4)
\end{equation}       %{\vskip-.1in} 
where $y(t) = f(t) + Z(t)$, with $Z(t)$ being a white noise process.
Away from the $m$-convexity change points, we linearize \ref{3.3C}
about $\fh^{(m)}(t) \sim f^{(m)}(t)$.
Let $\ftl(t)$ be the linearized variable for \ref{3.3C}:
$\ftl^{(m)}(t) \approx \fh^{(m)}(t) - f^{(m)}(t)$,
where $\ftl(t)$ satisfies
\begin{equation}\label{LIN}
(-1)^m (p-1) \lambda d^m [|f^{(m)} |^{p-2} \ftl^{(m)} (t)]  \ + \
\ftl(t) = Z(t)  \ .    
\end{equation}       %{\vskip-.1in} 
%where $\chi(t) \equiv {\rm sgn}[f^{(m)}(t)]$.      
When $\lambda |f^{(m)}(t) |^{p-2}$ is small but nonzero,
the homogeneous equation may be solved using the Wenzel-Kramer-Brillioun
expansion. The resulting Green's function for $\ftl$ may 
be recasted as a kernel smoother with an effective halfwidth: 
%For small $\lambda$, a two scale analysis is appropriate. 
%We make the ansatz that the limiting $\fh(t)$ 
%has the local kernel smoother form:
%\begin{equation}\label{K1} \fh^{(m)}(t) = \frac{1}{Nh^{m+1}(t)}
%\int_{-h(t)}^{h(t)} y(s) \kappa^{(m)}({t-s \over h(t)}) ds \ \ 
%\times \left[1 + \ {\cal O}(h) \right] \ , \end{equation}
%where $h(t)$ is the kernel halfwidth and $\kappa$ is a prescribed
%kernel. Since $h(t)$ is a small parameter, the solvability condition is
\BEQ \label{Heff}
h_{eff}(t) \sim [\lambda F'(t) |f^{(m)}(t)|^{p -2} ]^{1\over 2m}
\ .\NEQ
%In (\ref{Heff}), we have added a subscript $eff$ to denote the
%effective halfwidth.
%We assume that the spline estimate converges to a kernel smoother
%with halfwidth, $h_{eff}(t)$. 
%We replace $ d^m [|\fh^{(m)} |^{p-2} \fh^{(m)} (t) ] $ by 
%$(p-1) d^{m-1} [|f^{(m)} |^{p-2} \fh^{(m)} (t) ]$   
%on the assumptions that $|\fh^{(m)} (t) -\fh^{(m)} (t)|<< |\fh^{(m)}| (t)$
%and that the scale length of $\fh^{(m)} (t)$
%is much larger than that of $f^{(m)} (t)$.
%The resulting solution has an effective scale length of
%\BEQ \label{Heff}
%h_{eff} \sim [\lambda F'(t) |f^{(m)}(t)|^{p -2} ]^{1\over 2m}
%\ .\NEQ
%At present, our understanding of the statistics of false inflection
%points is limited to Gaussian errors and linear estimators. 
For $1\leq p \leq 2$, the effective halfwidth of the robustified
function  automatically reduces the halfwidth in regions of large
$|f^{(m)} (t)|$ just like a variable halfwidth smoother. 
We caution that this result has not been rigorously derived.

%Assuming that the robustified spline converges to a kernel smoother

%In practice, it is
%often advantageous to replace both the residual errors 
%and the penalty function with more robust analogs: \ 
%\BEQ
%$\sum_{i=1}^N |y_i -\hat{g} (t_i )|^{p} + \lambda \int
%|\hat{g}^{(m)} (t)|^{q_2} dt $, %\NEQ
%where $1\leq p \leq 2$. 
%Representation and duality theorems are given
%in [8] for $q_j > 1$.
%A heuristic scaling shows that the effective halfwidth of the
%robustified function satisfies $h_{eff} \sim [\lambda |f^{(m)}
%(t)|^{p -2} ]^{1\over 2m}$. 
For the equivalent kernel, the bias error scales as $f^{(m)} (t) h_{}^m$
while the
variance is proportional to $1/Nh_{}$. The halfwidth that
minimizes the mean square error scales as 
$h_{MSE} \sim \left[N|f^{(m)} |^2 \right]^{{-1\over 2m+1}}$
%, while the halfwidth to eliminate false change points of $g^{(\ell )}$ with
%asymptotic probability one satisfies $h_{cr} N^{{-1/ (2\ell +3)}}
%\rightarrow\infty$. The optimal variable halfwidth kernel smoother 
%has a kernel halfwidth proportional to $|f^{(m)} |^{{-2\over 2m+1}}$. 
The two halfwidths agree at $p =2/(2m+1)$, but $p<1$ is ill-conditioned.

%When $p =
%q_2 =1$, the problem reduces to a linear programming problem for each
%predetermined set of constraints.

We recognize that this derivation is formal, but we believe
that a rigorous multiple scale analysis may prove \ref{Heff}.
Our purpose is only to motivate the connection between robust splines
and adaptive halfwidth kernel smoothers.

\section{Constrained smoothing splines and duality} \label{DUAL}

In (\ref{3.3D}), the intervals on which $f^{(\ell )} (t)$ vanishes are unknowns 
and need
to be found as part of the optimization. Using the differential
characterization (\ref{3.1}) loses the convexity properties of the underlying
functional. For this reason, extremizing the dual functional is now
preferred.
%\end{enumerate}

\begin{thm}[Convex Duality]  \label{Dthm}
%\noindent{\bf Theorem 2.} 
{ 
%{\bf  P_{x}} g(t) = g(t)$ if $(-1)^{j-1}g^{(\ell )} (t) > 0$ 
%where $x_k \leq t < x_{j+1}$ and ${\bf P_x}g (t) = 0$
The dual variational problem of Theorem \ref{thm1} is: 
Minimize over $\alpha \in \RR^N$
%{\vskip-.13in}
\begin{equation}\label{3.4}
{\rm VP}^*[\alpha;{\bf x}] \equiv % \min
{\lambda^{1-q} \over q} \int |[{\bf P_x}^*B\alpha ]^{(m)} (s) |^q ds +
\sum_{i=1}^N \rho_i^* (\alpha_i ) - \alpha_i y_i \ ,
%\eqno (2.6)
\end{equation}
%{\vskip-.1in}
\noindent
where $\rho^*_i$ is the Fenchel/Legendre transform of $\rho_i$, and 
$B \alpha (t) \equiv \sum_i^N b_i (t) \alpha_i$ 
with $b_i(t) = L_i R(\cdot,t)$. 
The dual projection ${\bf P_x}^*$ is
defined as 
%the complement  of the projection onto dual cone of $V^{K,\ell}_{m,p}: 
%{\vskip-.1in}
\begin{equation}\label{3.5}
\int |[{\bf P_x}^* g]^{(m)} (s) |^q ds 
\equiv {\rm inf}_{\tilde{g}\in V^-}  \int_0^1 |g^{(m)}
-\tilde{g}^{(m)} (s)|^q\ ds
%{\rm\ s.t.\}  
\ , %\eqno(2.7)
\end{equation}
%{\vskip-.1in} 
%\noindent
%where the minimization is over $\tilde{g}$ in the dual cone subject to 
subject to the constraints $g^{(j)}(0)=\tilde{g}^{(j)}(0),\ \ 0\le j <m$.
The dual problem is strictly convex, %lower semi-continuous.
and its minimum is the negative of the infimum of (\ref{3.1}). 
When the $\{ h_i \}$ are linearly independent, the minimum satisfies
the differential conditions:
\begin{equation}\label{3.6D}
\alpha_i = \rho_i'\left(\langle h_i ,\fh \rangle -y_i \right) ,
\ \ {\rm and} \ \
\langle h_i ,f \rangle -y_i = \rho^{*\prime} (\alpha_i ) , \ \ \
\ \ i = 1\ldots N \ .\end{equation}
}
%{\vskip .05in}
\end{thm}

\noindent
{\bfab Proof.} Let $\chi_V$ be  the indicator function 
of $V^{K,\ell}_{m,p} [\xbf]$
%at $f$ 
and define
%{\vskip-.11in}
\begin{equation}\label{3.8}
U(f) = {\lambda \over p} \int_0^1 |f^{(m)} (s)|^p ds + \chi_V (f) \ .
\end{equation}
%{\vskip-.07in} \noindent
We claim that the Legendre transform of $U(f)$ is %the first term in 
(\ref{3.5}).
%where 
Note that $\chi_V^* (g) = \chi_{V^-} (g)$, the indicator function
of the dual cone $V^-$. The Legendre transform of the first term in 
(\ref{3.8})
is
%{\vskip-.15in} 
\begin{equation}%\label{}
V_1^* (g) = {\lambda^{1-q}\over q} \int_0^1 |g^{(m)} (s) |^q ds \ \ 
{\rm for}\ \ g \in W_{m,q}^0, \ \ \ {\rm and}\ \ \infty \ \ {\rm otherwise}.
\end{equation}               
\noindent
Our claim follows from % property of the Legendre transform is
%%{\vskip-.15in}
$[U_1 +U_2 ]^* (g) = \inf_{g'} \{ U_1^* (g-g' ) + U_2^* (g' ) \}$.
%$%{\vskip-.15in}
%where $P_v g(s) = 0$ for those $s$ where $g^{(\ell )} (s)$ violates the
%constraints and $P_v g = g$ otherwise.
\noindent
The remainder of the theorem including the differential conditions (\ref{3.6D})
follows from the general duality theorem of
Aubin and Ekeland \cite[ %Ch. 4.6, 
p.~221]{Aubin84}. 
\eopp

An alternative formulation of the duality result for quadratic smoothing
problems is given in \cite{MU91}. For both theories, the case $\ell < m$ 
is difficult to evaluate in practice because the minimization in
(\ref{3.5}) can only rarely be reduced to an explicit finite dimensional
problem. Only a few partial results are known 
when $\ell < m$ \cite{Deutsch,MU85,MU91}.
For the case $\ell=m$, the minimization over the dual cone can be done
explicitly and yields the following simplification:

\begin{cor}\label{mlcor}
%{\bf Corollary 3.} {\em 
{For $\ell=m$, the dual projection, ${\bf P_x}^*$,
is a local operator with  
$[{\bf P_x}^*M\alpha ]^{(m)} (s)$
$ = \sum_{i=1}^N \alpha_i b_i^{(m)}(t)$
if $\sum \alpha_i b_i^{(m)}(t) \chi_{\bf x}(t) \ge 0$ and zero otherwise.
Thus the minimization of (\ref{3.5}) is finite dimensional.
}\end{cor}

%obtaining explicit, finite dimensional representations
%of (\ref{3.5}) is problematic. Some 
%The negative polar of $V^{k,m }_{m,p} [\xbf]$ is
%$V^{k,m}_{m,p} [\xbf]^- = -V^{m,\ell }_{m,p} [\xbf] \cap W_{m,p}^0$. 
%For $\ell<m$, 
%The differential conditions for an extremum are precisely
%%{\vskip-.15in} \end{equation}%\label{}
%(-1)^m d^m [|f^{(m)} |^{p-2} f^{(m)} ] + M^* \alpha +N_k (f) = 0
%\end{equation}%\label{}%{\vskip-.15in}
%normal cone of $f$ is $N_{\ell} (f) = \{ g \in W_{m,p}|$
%$-g^{(\ell )} (t) \in V_{\ell ,m,p}$ and $g(t) \neq 0$ only if
%$f^{(\ell )} (t) = 0\}$. When $f^{(\ell )} (t) = 0$, an $a$ region,
%$M^* (t) \alpha_1 + N_{\ell} (f) (t)$ is zero. When $f^{(\ell )} (t) \neq
%$, $N_{\ell} (f) = 0$ and $(-1)^m d^m [|f^{(m)} |^{p-2} f^{(m)} t ]$
%and $M^* \alpha = 0$.%box 

%{\vskip-.12in}
\section{Change point estimation} \label{CPsect}
%{\vskip-.12in}
When the number of change points is fixed, but the locations are unknown, we
can estimate them by minimizing the functional in (\ref{2.3}) 
with respect to the
change point locations. We now show that there exists a set of minimizing
change points.

\begin{thm}\label{CPthm}
%{\bf Theorem 3.} 
{For each $K$ with $\ell=m$, there exist change points 
$\{ x_{k},\ k =1,
\ldots K\}$ which minimize the variational problem (\ref{2.3}).}
\end{thm}

\noindent
{\bfab Proof.} We use the dual variational problem (\ref{2.5}) 
and maximize over $\xbf \in [0,1]^K$ after minimizing over the 
$\alpha \in R^N$. We need only consider $\alpha$ in the compact region
$\sum_{i=1}^N \rho_i^* (\alpha_i ) - \alpha_i y_i $.
%$N+k$ dimensional space of $\{ \alpha ,x \}$. 
For $\ell=m$, explicit construction of
the functional (\ref{2.5}) shows that it is
jointly continuous in $\alpha$ and $\xbf$.  
Since  (\ref{2.5}) is convex in $\alpha$,
Theorem \ref{CPthm}
follows from the min-max theorem \cite[p.~296]{Aubin84}.
\eopp

We conjecture that Theorem \ref{CPthm} is true for $\ell <m$, but
we lack a proof that Eq.~(\ref{3.5}) is continuous with respect to
$\xbf$ for $\ell <m$.
The change point locations need not be unique. The proof requires $\le$ 
instead of $<$ in the ordering
$x_k \leq x_{k+1}$ to make the change point space compact. When
$x_k = x_{k+1}$, the number of effective change points is less than $K$.

%In \cite{Riedel2}, %Section \ref{CLsect}, 
%we show that under certain conditions
%the locations of the empirical change points converge to the actual
%change points.

In \cite{Mammen91}, Mammen considers the case where $K$ is known
but the locations are unknown. The function is estimated 
using simple least squares on a class of functions roughly analogous
to $V^{K,m+1}_{m,\infty}$.
% \cap \{${\em functions with a prescribed H\"older bound on } $f^{(m)} \}$. 
Mammen proves that this estimate achieves
the optimal convergence rate of $N^{-2m/(2m+1) }$ for the 
mean integrated square error. %ISE.
Unfortunately, Mammen's estimator is not unique and often results in
aestetically unappealing fits.
%this infinite dimensional optimization cannot be
%reduced to a finite dimensional problem. Therefore, the estimator
%in \cite{Mammen91} is only a theoretical construct and not a 
%practical algorithm.

For both formulations.
finding the  optimal change points locations % , $\xbf$, 
is computationally intensive. For each candidate set of change points,
the the likelihood function needs to be minimized subject to PC constraints.
One advantage of our formulation is that for each candidate value of $\xbf$,
the programming problem is strictly convex in the dual. 
This strict convexity is lost if one uses a penalty functional with 
$p= \infty$ as in \cite{Mammen91} or $p=1$ corresponding to a total
variation norm. If the total variation norm is used and an absolute
value penalty function is employed ($\rho(y) = |y|$), the programming
problem reduces to  constrained linear programming.
%and requires the solution of a convex programming problem 

\section{Discussion}

We have considered robust smoothing splines under piecewise convex 
constraints. We generalize the standard representation and smoothness
results to nonlinear splines using convex analysis. 
When the same derivative is both constrained and penalized ($\ell=m$),
the dual problem is finite dimensional.

We have sketched a derivation of the effective halfwidth
of a robust spline.
By robustifying the functional, the effective halfwidth (\ref{Heff})
for the equivalent kernel smoother
scales as $|f^{(m)}(t)|^{(p-2)/2m}$.
The halfwidth that
minimizes the mean square error scales as 
$h_{MSE} \sim \left[N|f^{(m)} |^2 \right]^{{-1\over 2m+1}}$.
Thus robust splines adjust the halfwidth, but not as much as 
the asymptotically optimal local halfwidth would.
%comes the closest to $h_{MISE} to make the asymptotically
%while the halfwidth to eliminate false change points of $g^{(\ell )}$ with
%asymptotic probability one satisfies $h_{cr} N^{{-1/ (2\ell +3)}}
%\rightarrow\infty$. The optimal variable halfwidth kernel smoother 
%has a kernel halfwidth proportional to $|f^{(m)} |^{{-2/(2m+1)}}$. 
%The two halfwidths agree at $p =2/(2m+1)$, but $p<1$ is ill-conditioned.
When the number of convexity change points is known, their locations
may be estimated by minimizing the penalized likelihood. For
$\ell=m$, we have existence, but not necessarily uniqueness.

%The advantage of the pilot es

%\begin{enumerate}
{\bf Acknowledgments:}
{
Work funded by U.S. Dept.\ of Energy Grant DE-FG02-86ER53223.}
%{\vskip-.1in}

%\References

%\begin{center}
%REFERENCES
%\end{center}

%\end

\end{document}